\input{aipcheck}

\documentclass[
    final            % use final for the camera ready runs
  ]
  {aipproc}

\layoutstyle{8x11double}

\begin{document}

\title{Sites for Gamma-ray Astronomy in Argentina}


\author{A.C. Rovero}{
  address={Instituto de Astronom\'ia y F\'isica del Espacio (CONICET-UBA), CC 67, 1428,
  Cdad. de Buenos Aires, Argentina},
}
\author{G.E. Romero}{
  address={Instituto Argentino de Radioastronom\'ia (CCT-LP, CONICET), CC 5, 1894,
  Villa Elisa, Buenos Aires, Argentina}
}
\author{I. Allekotte}{
  address={Centro At\'omico Bariloche and Instituto Balseiro (CNEA and UNCuyo), 8400,
  Bariloche, Rio Negro, Argentina}
}
\author{X. Bertou}{
  address={Centro At\'omico Bariloche and Instituto Balseiro (CNEA and UNCuyo), 8400,
    Bariloche, Rio Negro, Argentina}
}
\author{E. Colombo}{
  address={Laboratorio Tandar (CNEA), Av. Gral. Paz y Constituyentes, Buenos Aires, Argentina}
}
\author{A.~Etchegoyen}{
  address={Laboratorio Tandar (CNEA), Av. Gral. Paz y Constituyentes, Buenos Aires, Argentina}
}
\author{B. Garcia}{
  address={Universidad Tecnol\'ogica Nacional, Facultad Regional Mendoza, Rodr\'iguez 273,
  5500, Mendoza, Argentina}
}
\author{D. Garcia-Lambas}{
  address={Instituto de Astronom\'ia Te\'orica y Experimental (CONICET-UNC),  
  Laprida 854, 5000, C\'ordoba, Argentina}
}
\author{H. Levato}{
  address={Complejo Astron\'omico El Leoncito, CC 467, 5400, San Juan, Argentina}
}
\author{M.C. Medina}{
  address={Laboratoire de l'Univers et ses Th\'eories, Obs. de Paris,
  5 Place Jules Janssen, 92195, Meudon Cedex, France}
}
\author{H.~Muriel}{
  address={Instituto de Astronom\'ia Te\'orica y Experimental (CONICET-UNC),  
  Laprida 854, 5000, C\'ordoba, Argentina}
}
\author{P. Recabarren}{
  address={Instituto de Astronom\'ia Te\'orica y Experimental (CONICET-UNC),  
  Laprida 854, 5000, C\'ordoba, Argentina}
}

\begin{abstract}
We have searched for possible sites in Argentina for the installation of large air 
Cherenkov telescope arrays and water Cherenkov systems. At present seven candidates
are identified at altitudes from 2500 to 4500 m. The highest
sites are located at the NW of the country, in La Puna. Sites at
2500 and 3100 m are located in the West at El Leoncito Observatory,
with excellent infrastructure. A description of these candidate sites is
presented with emphasis on infrastructure and climatology.
\end{abstract}

\maketitle

%%%%%%%%%%%%%%%%%%%%%%%%%%%%%%%%%%%%%%%%%%%%
%% MAINMATTER
%%%%%%%%%%%%%%%%%%%%%%%%%%%%%%%%%%%%%%%%%%%%

\section{Introduction}

After successful operation of Cherenkov telescopes, like HESS, MAGIC and VERITAS,
the scientific community is beginning to
seek for sites to locate a new generation of instruments. 
Not only air Cherenkov but 
also water Cherenkov detectors are expanding frontiers. HAWC
is being constructed in Mexico and would require a counterpart in the
Southern Hemisphere. At smaller scale, LAGO has been expanding from Auger,
in Argentina, to Bolivia, Mexico, Venezuela and Per\'u, seeking for high altitude
sites to install single water Cherenkov tanks.

For galactic gamma-ray astronomy the observation from a Southern Hemisphere
location has obvious advantages as the most important portion of the Galactic
Plane is seen from the South, including the Galactic Center. For extragalactic
gamma-ray astronomy, a homogeneous distribution of instruments should be
spread over the World, especially for monitoring of transient sources,
like AGNs, which require a complete time coverage (RA coverage). In the
Northern Hemisphere the distribution of air Cherenkov telescopes is such
that this could be achieved. The lack of telescopes in South America, however,
prevents the correct time coverage for the Southern Hemisphere. For water
Cherenkov detectors 
this problem is attenuated by their large angular acceptance.

Requirements for gamma ray detectors are wide. For water Cherenkov, 
high altitude and water availability have to be considered. For air Cherenkov,
moderate altitudes, atmospheric transparency and low cloud coverage are required.
In any case, good infrastructure and local support are desired.
In South America moderate and high altitude sites are located in
"La Puna" and its surroundings. This region extends from the NW of
Argentina, up to the North of Chile, West of Bolivia and SW of Per\'u.
In this area the cloud cover improves toward the SW so only the North
of Chile and NW of Argentina would be available for Cherenkov telescopes.
In Chile there are several Observatories with optical, infrared and millimetric
telescopes. The ALMA array is being deployed and might extend its aperture by
locating a single dish in Argentina \cite{mirabel}. On the East of La Puna, in
Argentina, there are sites being characterized for Extremely Large Telescopes
(ELT) by a local group (for ESO). Optical (CASLEO: www.casleo.gov.ar), Radio
(IAR: www.iar.unlp.edu.ar) and cosmic-ray (Auger: www.auger.org) observatories
are operating in the country, with an important 
academic community.

In the last two years we have been searching for sites in Argentina that fulfill
conditions for both water and air Cherenkov detectors. Here we report on 
such searches.

\section{Explored regions}
\begin{figure}[!ht]
\centering
\hbox{
 \includegraphics[height=.27\textheight]{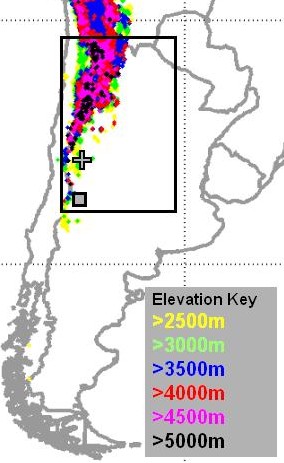}
 \includegraphics[height=.27\textheight]{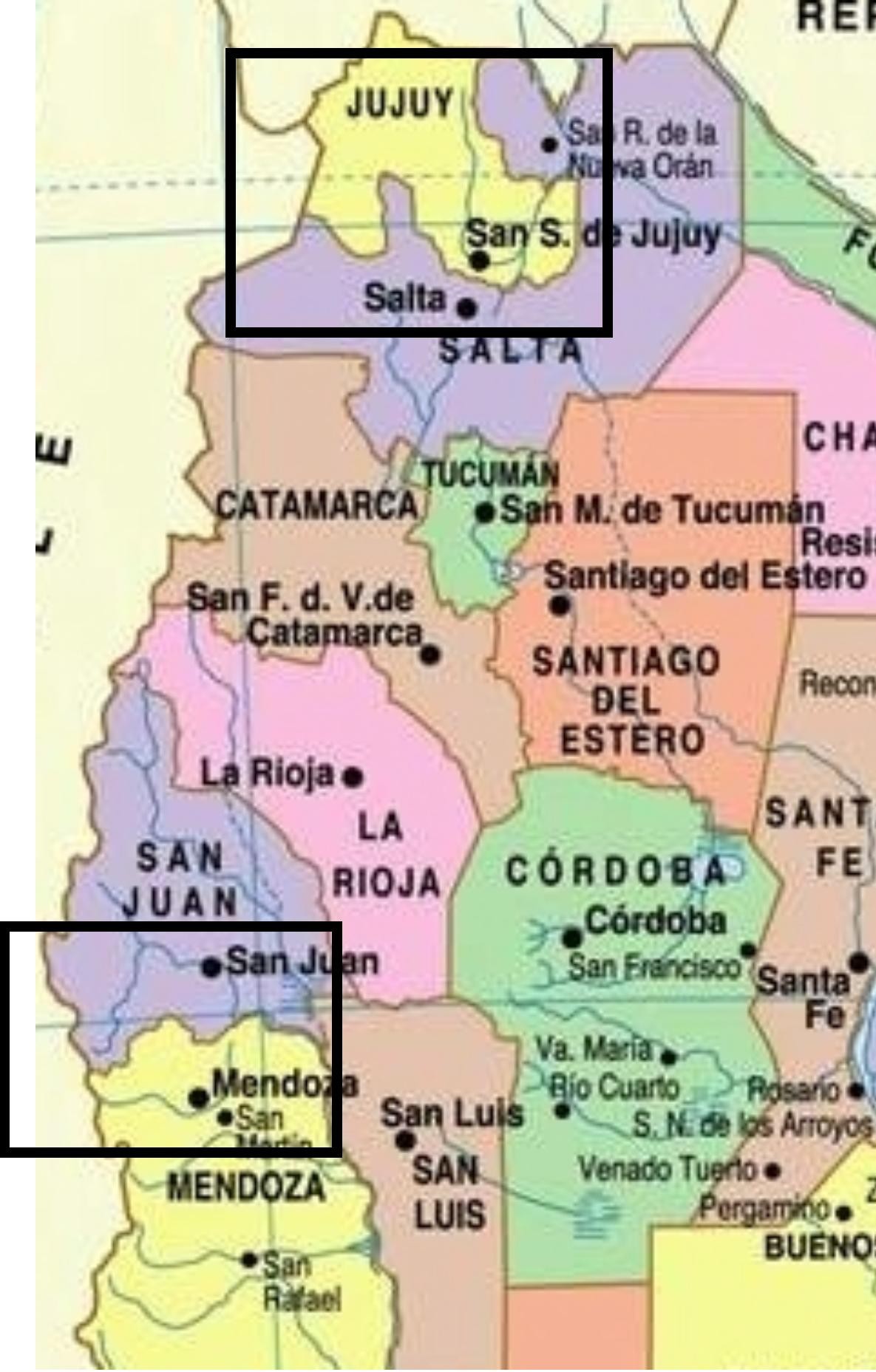}
      }
\caption{\footnotesize {Explored regions in Argentina. Left: Search for 1 km$^2$
sites \cite{fegan}. Square: Auger Southern Observatory (1400 masl);
cross: former candidate site for SKA (2600 masl), 12 km North of El Leoncito Astronomical
Observatory. The enclosed area is enlarged on the right. Right: province
limits in the Center and NW of the country. The explored region enlarged in
Figures \ref{acaNOA} and \ref{acaCASLEO} are indicated.}}
\label{niveles}
\end{figure}

To seek for a flat land of $\sim$1 km$^2$ at more than 2500 meters above sea level (masl)
with clear skies in Argentina, the Central and NW regions of the country have to be
considered. In Figure \ref{niveles} a search for 1 km$^2$ sites at different altitudes
is shown \cite{fegan}. So far we have searched
two subregions, one at high altitudes on the North and another at moderate 
altitudes on the South (see Figure \ref{niveles}, right). The
northern region is part of La Puna, a land at more than 3500 masl. In Argentina this
region is located to the East of the Andes mountains, on the provinces of
Jujuy, West of Salta, and North of Catamarca.
The region at moderate altitudes extends to the South of La Puna, on the West
of the provinces of La Rioja, San Juan and Mendoza. In the following sections
we report on two searched regions, the northern portion of La Puna and 
the National Reserve of ``El Leoncito'', in San Juan.

\section{High altitude sites: La Puna}

\begin{figure}[!ht]
  \includegraphics[height=.25\textheight]{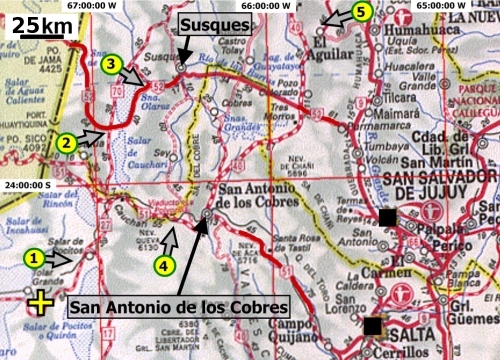}
  \caption{\footnotesize{Road map of the northern area indicated in Figure
\ref{niveles}. Arrows point the locations selected as candidate sites:
(1): Salar de Pocitos; (2): El Porvenir; (3): El Taire; (4): Alto Tocomar
and (5): Mina Aguilar. The cross shows the chosen position for ELT being
characterized by the IATE group: Tolar Grande. Squares indicate the position
of nearby cities with daily flight connections to Buenos Aires.}}
  \label{acaNOA}
\end{figure}

A map of the northern explored region is shown in Figure \ref{acaNOA}.
La Puna is climatologically classified as a highland climate. It
has low temperatures, poor vegetation, and a low annual
precipitation with a dry winter. It extends along the leeward
slopes of the Andes where winds are dominated by the Pacific high
pressure system (subsidence). Advection from the Pacific causes a
predominant wind from the West which is the main control of
climate in the whole Puna \cite{sch}. A roughly similar West-East regime
exists for rainfall, humidity and cloudiness.

As an example, we can cite meteorological data from two stations. 
One is near Susques, at 3675 masl, and the
second is La Casualidad (25:03 S, 68:13 W), located at 4092 masl, 300 km
SW of Susques. From Susques we have data from 1972 to 1990
(Direcci\'on de Hidr\'aulica de Jujuy), only on precipitation and temperature:
average annual total rainfall is 184 mm (mainly in summer) and average
temperature is 7.6 Celsius. From La Casualidad we have ten years of data
from 1963 to 1972 \cite{servicio}: annual averages are: wind: 26\% calm,
55\% from West (28 km/h), 13\% from NW (22 km/h) and 5\% from
SW (24 km/h); Sky coverage: 22 days; humidity 33\%; temp.: 3.9
Celsius; low temp.: - 4.1 Celsius; frost: 267 days; total rainfall: 37 mm in 6.3
days; snow: 2.8 days.

By analyzing historical climate data, Schwerdtfeger \cite{sch} shows that
cloudiness is another variable that seems to show roughly the same
West-East behaviour in La Puna. Dedicated studies have been done by
the IATE group in their search for ELT \cite{recabarren}
and by Erasmus \cite{erasmus} for the ALMA project using satellite images. 
Results from these analyses are all coherent with each other. The 
result on sky cover shown by Schwerdtfeger indicates that the isoneph
of 20\% is right to the East of the whole region studied by both Erasmus 
and IATE. All of them indicate that cloudiness decreases toward the West. 
For the selected sites in Figure \ref{acaNOA} we note that the sky cover conditions
are similar in average ($\approx$~85\% of clear nights), except for Mina Aguilar (5)
which is more into the cloudy skies. However, if a seasonal
analysis is done, the northern sites, (2) and (3), have larger variations than
the other two locations in the South, (1) and (4), being better in winter
and worse in summer.

This region has two major cities with international airports and daily
flights to Buenos Aires. {\it San Salvador de Jujuy} (pop. 250,000)
and {\it Salta} (pop. 500,000). They host their national universities:
{\it U. de Jujuy} ($\approx$10,000 students) and
{\it U. de Salta} ($\approx$20,000 students). Two towns are
identified in this area, {\it Susques} and {\it S.A. de
los Cobres}.

Susques (23:23:12 S, 66:21:47 W) is a small town (pop. 2000)
at 3675 masl. It takes 2.5-3 hours ride from
Jujuy city (202 km) driving on a very nice paved road
that connects to the West with Antofagasta (Chile), through the
border point ``Paso de Jama'' (4400 masl). There are two good hotels in
Susques both located on the National Route 52 at 3700 masl: ``Hotel
Unquillar'' (www.elunquillar.com.ar) and ``Hotel Pastos Chicos''
(www.pastoschicos.com.ar), with an overall allocation of 60 beds in 20 rooms
at $\approx$~u\$s30 (single rate). Satellite TV and regular and cell
phone services are available.

San Antonio de los Cobres (SAC) (24:13:30 S, 66:19:10 W) (pop. 3000) is 
at 3775 masl. From Salta city it takes 2.5 hours (164 km) driving on the National
Route 51, which is 75\% of the way a paved road. Route 51 continues as gravel road
to the West to connect with Chile at ``Paso Sico'' (4092 masl).
There is one good hotel in town, ``Hotel de las Nubes'', which has
approximately 25 beds in 10 rooms at a similar rate of Susques.
Satellite TV and regular and cell phone services are available.

\subsection{Identified candidate sites}
\label{punaSites}

{\it Tolar Grande:} Astronomers from IATE are characterizing this
site for the location of ELT (ESO). Their chosen
point is in the Mac\'on mountain range, 25km East of Tolar Grande (24:38:49 S, 67:20:14 W)
at 4600 masl. There are several experiments being carried out there to
characterize the sky.
This led us to consider Tolar Grande as a reference site, where
flat lands at $\sim$4000 masl can be found. It takes 3.5-4 hours to get there
from SAC and there is a nice lodging house at low rates.

{\it Salar de Pocitos:} 80 km SW of SAC we have identified a candidate
site nearby the small neighborhood of Salar de Pocitos, where no significant
infrastructure is available except electricity, water and phone connection.
It takes 2-2.5 hours drive from SAC (140 km) on a very nice gravel road
to get there. The site is a very extended plain at 3750 masl
(slope $\approx$0.01), limiting with a mountain range to the West and a salt land
to the East, 5 km West of route 27 (24:26:40 S, 67:06:10 W).
Ground is compact and
vegetation is very poor. There is a railroad crossing the field. The salt land
on the East should not contribute with aerosols as wind is mainly from the West.

{\it El Porvenir:} Few km South from the intersection of routes 52 and 70, we
identified a possible site at 3950 masl (23:38:30 S, 66:46:30 W). From there, it
takes 40 minutes drive (65 km) on the paved route 52 to get Susques, where a base of
operation could be located. There are $\approx$5 km of flat plain from route 70 before
reaching the mountain range on the West. There is also a salt land to the East which
should not contribute to the air aerosol content as wind is mainly from the West.

{\it El Taire:} This is a site specially identified for water Cherenkov systems
(23:23:41 S, 66:31:27 W). It is at 4150 masl closeby route 52, 20 km West of Susques.
There is a stream of spring water and perforation is a general practice to get
water from underground.

{\it Alto Tocomar:} This is the highest plain we have identified (24:12:00 S; 66:30:30 W).
It is 25 km West of SAC at 4500 masl, closeby route 51, with nearby summits of more
than 5000 masl and several km$^2$ of flat land. There is a railroad crossing the site
and few spring water points have been localized at the western border. 

{\it Mina Aguilar:} Another excellent site for water Cherenkov systems
(23:09:03 S, 65:42:55 W). It is at 3900 masl, nearby a small town with all supplies
needed for operations, including stores and hospital. A site at 5100 masl is easy
to access by vehicle in 20 min. drive.

\section{Moderate altitude Sites}

\begin{figure}[!ht]
  \includegraphics[height=.23\textheight]{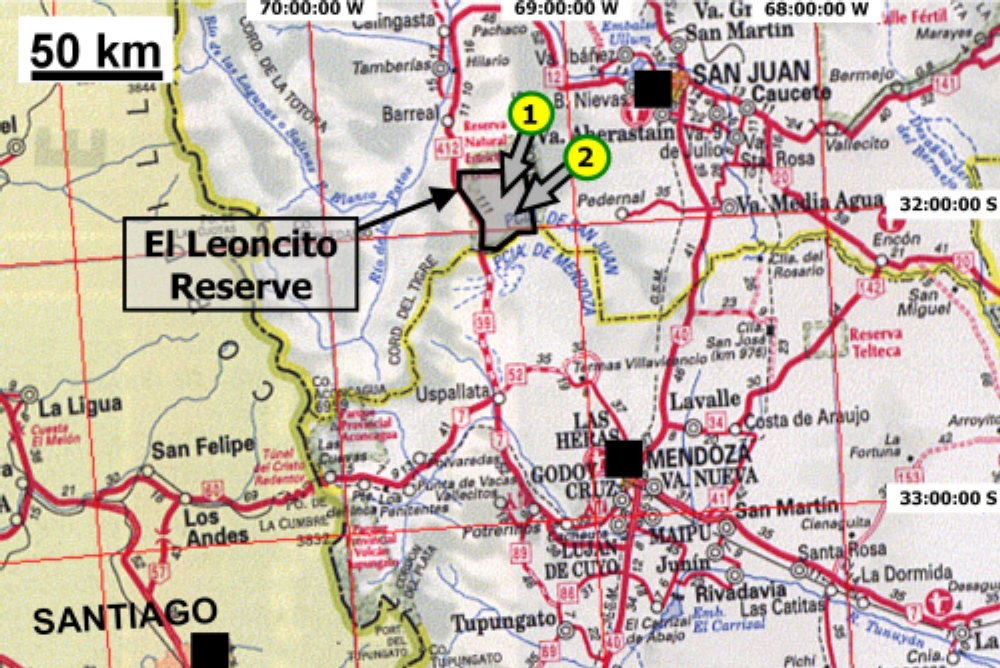}
  \caption{\footnotesize{Road map of southern area indicated in Figure \ref{niveles}.
Arrows point candidate site locations: (1): Naranjo; (2):
Tontal. Squares: position of nearby cities with daily flights to Buenos Aires.}}
  \label{acaCASLEO}
\end{figure}

{\it El Leoncito Astronomical Observatory} (CASLEO: Complejo AStron\'omico el LEOncito)
(31:48:00 S, 69:17:30 W) is an astronomical facility located at 2600 masl in the
province of San Juan, with 
offices and laboratories in the city of San Juan.
In Figure \ref{acaCASLEO} we show the road map of the area. The Observatory is placed
inside a reservation area of 416 has. dedicated to astronomy, which is part of
the National Park ``El Leoncito'', a 70,000 has. natural reserve. There are two major
cities with daily connection to Buenos Aires, {\it San Juan} (pop. 450,000), and
{\it Mendoza} (pop. 900,000). These two cities are the host of their national
universities: {\it U. de San Juan} ($\approx$20,000 students),
and {\it U. de Cuyo} ($\approx$38,000 students).
There is a small town, Barreal (pop. 2500), 40 km NW (30 min. drive)
of the Observatory where basic supplies are available. 
Regular transportation to the mountain is organized by the Observatory staff
several times per week. It takes 3.5 hours to get from San Juan to the
Observatory (250 km) on a good paved road, through Calingasta and Barreal. One
alternative for international travelers is to flight to Santiago (Chile), connect
to Mendoza and take a rented car to reach the site through Uspallata.
It also takes 3.5 hours to get from Mendoza city to the Observatory (230 km),
mostly on paved road.

The most distinctive feature of a site in this area is the Observatory infrastructure.
After more than 20 years of operation the site has everything needed for operating and
maintaining any running instrument on the mountain.
A 13.2 KV power line, an emergency generator, internet (25 Mbps) and 
regular and cell phone services are available.
There are on the mountain mechanical and electronics workshops, a liquid nitrogen plant 
and a 2.15 m mirror coating chamber. The lodging facilities include bedrooms with
private toilets for 45 people and a dining room for the same number.
More than ten people from the staff are permanently at the site for maintenance,
cleaning, cooking and to attend visitors.

The Observatory has its own meteorological station
(www.casleo.gov.ar/weather/leonci\_weather.htm). Monthly values for the last
ten years are: mean temperature: 5.5 C (July) and 17.5 C (January); maximum
temperature: 20.3 C (June) and 30.5 C (January); rel. hum.: 27-43 \% along the
year; rainfall: 11-18 mm (December-March) and 2-7 mm (April-November); 
wind speed: 4.5-8.4 km/hr, mainly from SW (55~\%) and from SE (30~\%); maximum
wind speed: 62-88 km/hr along the year; hail: 1-2 days/yr; frost: none; lightning:
6-8 days/yr; maximum snowfall: <30 cm.

Analyzing more than 20 years
of operations the Observatory has 270 observing nights per year (74~\%), from
which 100 nights are photometric (more than 6 hr of perfect sky conditions). 
Sky brightness has been measured in magnitudes per square arcsecond: U~=~22.1,
B~=~23.3 and V~=~22.7. For comparison, these values are similar to those
at Cerro Tololo Inter-American Observatory (from CTIO), Chile: U~=~22.0,
B~=~22.7 and V~=~21.8.

In an attempt to connect both the northern and southern studied regions,
the IATE group analyzed 1.5 years of GOES images (1998-1999) of both El Leoncito
and La Puna. The uncertainties of
this study are unknown but good enough to say that between site candidates on the
North and South there is a difference of $\approx$~10\% in sky coverage, being
the northern sites better and characterizing El Leoncito with $\approx$~75\% of clear
nights.

\subsection{Identified candidate sites at El Leoncito}

{\it Tontal:} To the West of the Tontal range there is a plain of at least 5$\times$10 km$^2$ at
3000 to 3500 masl. We have pointed to a specific site at 3130 masl (31:48:12 S; 69:10:31 W),
located $\approx$~15~km to the East of the Observatory. Once in the site, accessibility is very good.

{\it Naranjo:} To the North of the Observatory on the West of El Naranjo range there is a site
of similar characteristics of Tontal but with slightly smaller slope.
The marked location is at 2590 masl (31:43:52 S; 69:16:20 W). This site was the former
SKA candidate site. It is very easy to access from the Observatory by a 20 min. drive (10 km).
Ground is compact assuring accessibility.

\section{Final Comments}

We have searched for candidate sites for ground-based gamma-ray astronomy
in two regions of Argentina: a northern region, in La Puna, and a southern
one, nearby the El Leoncito Astronomical Observatory. In La Puna we have
found five candidate sites at altitudes from 3750 to 4500 masl, being the ones
on the SW better suited for air Cherenkov detection, 
as there are $\approx$~85~\% of clear nights.
Sites near the town of Susques (El Taire and El Porvenir) have very good 
accessibility and reasonable infrastructure. The highest site (Alto Tocomar)
is 25 km from San Antonio de los Cobres, where a base for operations can be
mounted. The lowest site in the area, Salar de Pocitos, is the best candidate
for air Cherenkov although infrastructure is rather poor. The northeastern
site, Mina Aguilar, has all needed for water Cherenkov systems but is too
far into cloudy skies for air Cherenkov detectors. Earthquake risk is moderate
in La Puna. At El Leoncito we identified two sites at 2590 and 3130 masl.
El Leoncito Observatory has been operational for 20 years and has excellent 
infrastructure including lodging, power, internet,  phone, food, mechanical/electronics  
workshops, regular transportation, etc. It has
$\approx$~75~\% of clear nights. Earthquake risk is of some importance but
similar to other astronomical observatories.

\vspace{.4cm}
\noindent{\bf Acknowledgments:} We are grateful to Olga Penalba for information on
climatology and the local SKA group for site
information. The work at IATE is being supported by ESO.
The following authors are members of CONICET: ACR, GER, XB, AE, BG, DGL,
HL, HM and PR.



\bibliographystyle{aipproc}   % if natbib is available

%\bibliography{sample}

\IfFileExists{\jobname.bbl}{}
 {\typeout{}
  \typeout{******************************************}
  \typeout{** Please run "bibtex \jobname" to optain}
  \typeout{** the bibliography and then re-run LaTeX}
  \typeout{** twice to fix the references!}
  \typeout{******************************************}
  \typeout{}
 }

%%%%%%%%%%%%%%%%%%%%%%%%%%%%%%%%%%%%%%%%%%%
%% The following lines show an example how to produce a bibliography
%% without the help of the BibTeX program. This could be used instead
%% of the above.
%%%%%%%%%%%%%%%%%%%%%%%%%%%%%%%%%%%%%%%%%%%

\end{document}